\documentclass{iau} 
\usepackage{graphicx}

\title[IAUS247. Early Science with ELTs] %% give here short title %%
{ELT Spectroscopy of Extragalactic\\Massive Stars}

\author[C.~J.~Evans et al.]   %% give here short author list %%
{C.~J.~Evans$^1$, O.~H.~Ram\'{i}rez-Agudelo$^1$, M. Garcia$^2$,\\M. Puech$^3$ \& Y. Yang$^3$}

\affiliation{\begin{flushleft}
$^1$\,UK ATC, Royal Observatory, Blackford Hill, 
Edinburgh, EH9 3HJ, UK \\$\phantom{^1}$\,email: {\tt chris.evans@stfc.ac.uk} 
\\[\affilskip]
$^2$\,CAB, CSIC-INTA, Ctra. de Torrej\'{o}n a Ajalvir km-4, 28850 Torrej\'{o}n de Ardoz, Madrid, Spain\\
$^3$\,GEPI, Obs. de Paris, PSL University, CNRS, 
5 Place Jules Janssen, 92190 Meudon, France

\end{flushleft}}

\pubyear{2018}
\volume{247}  %% insert here IAU Symposium No.
\jname{Early science with ELTs}
\editors{N. Przybilla, K. Pollard \& A. Calamida, eds.}
\begin{document}

\maketitle

\begin{abstract}
  Quantitative spectroscopy of O-type stars in galaxies beyond 1\,Mpc
  requires the unprecedented sensitivity of the so-called Extremely
  Large Telescopes.  Visible spectroscopy with these exciting new
  facilities will give us the first chance to explore the properties
  of large samples of massive stars in very metal-poor galaxies, as
  well as the broad range of environments in the galaxies of the
  Sculptor Group and beyond. In this context, we introduce performance
  simulations of visible spectroscopy of extragalactic massive stars
  with the MOSAIC instrument, which is a multi-object spectrograph in
  development for the European Extremely Large Telescope (ELT).
  Preliminary results demonstrate that we can obtain sufficient
  signal-to-noise from one night of observations in galaxies out to
  distances of 3 to 4\,Mpc ($V$\,$=$\,\,22-23\,mag). We also highlight
  the goal of obtaining ultraviolet spectroscopy of the same
  individual stars beyond 1\,Mpc; this will require a future
  large-aperture space telescope with UV capabilities such as the
  LUVOIR concept currently under study.

\end{abstract}

\firstsection % if your document starts with a section,

\section{Introduction}
Spectroscopic surveys in the Milky Way and Magellanic Clouds have
provided a wealth of empirical data to improve and refine our
understanding of the physics and evolution of massive stars (e.g.
Evans et al. 2005, 2011; Sim\'{o}n-D\'{i}az \& Herrero, 2014).
Significant breakthroughs have included: validation of the predicted
metallicity ($Z$) dependence of the intensity of stellar winds (Mokiem
et al. 2007), recognition and characterisation of the prevalence of
binarity (multiplicity) in massive stars (Sana et al. 2012, 2013), and
identification and assessment of the role of very massive stars
(M\,$>$\,150\,M$_\odot$; Crowther et al. 2010, 2016). Improving our
understanding of the evolutionary pathways of massive stars has been
given further impetus by the GW170817 gravitational-wave detection of a
neutron-star merger and its resulting kilonova (e.g. Pian et al.
2017; Smartt et al. 2017).

A long-standing goal has been to extend detailed spectroscopic studies
of massive-star populations to metallicities lower than that of the
Small Magellanic Cloud (with $Z$\,$\sim$\,0.2\,$Z_\odot$) and to test
our latest models in a wider range of galaxies (e.g. those of the
Sculptor Group). However, in terms of distance, we have reached the
limit of what is possible with current large ground-based telescopes,
with observations of O-type stars out to $\sim$1.2 Mpc (e.g. Tramper et al.
2011, 2014, Garcia \& Herrero, 2013, Camacho et al. 2016). Even then,
we are limited to the most luminous (partially evolved) stars --
visible spectroscopy of main-sequence massive stars beyond the Local
Group requires the unprecedented sensitivity of the ELTs.

\section{Visible spectroscopy with MOSAIC}
The wider science case and design of the MOSAIC instrument is reviewed
in more detail by Morris et al. (these proceedings). In brief, the
Phase A design study of MOSAIC was completed in early 2018 
and provides two observational modes:
\begin{itemize}
\item{{\it High multiplex:} Integrated-light observations of up to 200
  objects at the angular resolution delivered by ground-layer adaptive
  optics (GLAO).}
\item{{\it High definition:} Observations of multiple sub-fields
    across the science focal plane at much finer angular resolution,
    provided by multi-object adaptive optics (MOAO).}
\end{itemize}
\smallskip The high-multiplex mode can be used to observe targets with
either visible or infrared spectrographs. The design of the visible
spectrographs at the end of Phase~A provides the spectral coverage and
resolving powers summarised in Table~\ref{mosaic_vis}. The blueward
extent of the visible spectrographs for MOSAIC was a point of detailed
analysis during the Phase~A study. In short, the necessary coatings of
the ELT mirrors strongly impact the efficiency in the blue visible,
and the Band 1 range in the current design only extends down to
4600\,\AA. In the context of spectroscopy of massive stars this means
we miss the H$\gamma$ line, an important diagnostic of stellar gravity
(cf. the more wind-influced H$\beta$ and H$\alpha$ lines). We will
revisit the blueward cut-off (and the parameters of the
high-resolution settings) in the next phase of the project --
extension to $\sim$4300\,\AA, even at diminished efficiency, would
still be valuable.

\begin{table}[h]
\begin{center}
\caption{Coverage and resolving power of MOSAIC visible spectrographs at end of Phase~A.}\label{mosaic_vis}
{\scriptsize \smallskip
\begin{tabular}{|l|c|c|}\hline
{\bf Setting} & {\bf $\lambda$-range [\AA]} & {\bf $R$} \\\hline
Band 1    & $\phantom{1}$$\phantom{1}$$\phantom{1}$4600-5840$\phantom{1}$$\phantom{1}$$\phantom{1}$ & 
$\phantom{1}$$\phantom{1}$$\phantom{1}$$\phantom{1}$5\,000$\phantom{1}$$\phantom{1}$$\phantom{1}$\\
Band 2    & 5700-7220 & $\phantom{1}$5\,000 \\
Band 3    & 7030-8900 & $\phantom{1}$5\,000 \\
Band 2 HR & 6360-6760 & 16\,500 \\
Band 3 HR$\phantom{1}$$\phantom{1}$$\phantom{1}$ & 8400-8850 & 17\,000 \\
\hline
\end{tabular}}
\end{center}
\end{table}

\subsection{Simulated MOSAIC observations}
To investigate the performance of MOSAIC for observations of
extragalactic massive stars we used the {\sc websim} tool developed by
Puech et al. (2016). {\sc websim} provides end-to-end simulations of
ELT observations, taking into account all of the relevant instrument
and telescope parameters (aperture, throughput, AO-corrected PSF,
detector properties, etc.). To simulate MOSAIC observations the user
inputs a template science spectrum, and {\sc websim} outputs a
simulated FITS frame (or datacube) to enable analysis with standard
tools. Our simulations used an input spectrum calculated with the 
{\sc fastwind} model-atmosphere code (Puls et al. 2005), with physical
parameters consistent with those expected for a mid O-type giant
($T_{\rm eff}$\,$=$\,35\,kK, log\,$g$\,$=$\,3.5,
$v$sin$i$\,$=$\,200\,km\,s$^{-1}$). 

\begin{figure}[]
\begin{center}
\includegraphics[height=7.75cm]{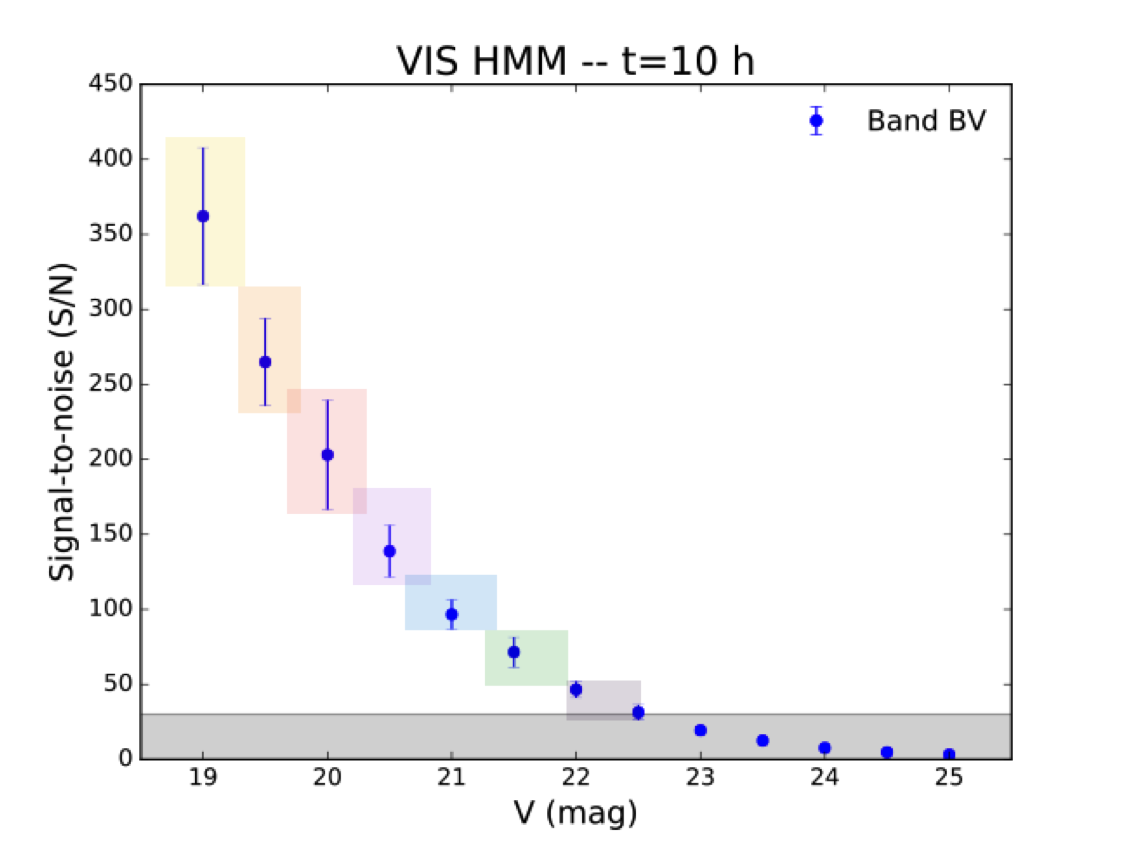} 
\caption{Mean signal-to-noise ratios obtained from simulated visible
 spectroscopy of extragalactic massive stars with MOSAIC.}
\label{fig1}
\end{center}
\end{figure}

To investigate the distance to which we will be able to recover
sufficient signal-to-noise (S/N) for quantitative analysis, we
simulated a range of magnitudes (19\,$<$\,$V$\,$<$\,25), with ten runs
of {\sc websim} for each. The mean continuum S/N (per \AA, roughly
equivalent to the resolution) from the runs for Band~1
($\lambda$4600-5840\,\AA) are shown in Fig.~\ref{fig1}. These results
are very promising for observations of the H and He I/II lines in this
region, and we are now simulating observations of the Band~2 region
(which includes H$\alpha$).

In brief, a 10\,hr integration delivers S/N\,$>$\,50 down to
$V$\,$\sim$\,22.5\,mag. This will open-up spectroscopy of mid-to-late
O-type dwarfs out to $\sim$3\,Mpc, and bright giants/supergiants out
to 5-6\,Mpc for the first time. In addition to accessing the large
spirals of the Sculptor Group (and beyond) this will give access to
main-sequence massive stars in the galaxies of the NGC\,3109 assocation
(e.g. Bellazzini et al. 2013). In the closer systems we will not need
such long integrations, enabling their populations to be characterised
in a small number of nights (and enabling multi-epoch observations to
investigate binarity).

\section{Future ultraviolet missions}

MOSAIC and the ELT will be a powerful tool with which to investigate
the massive-star populations in galaxies beyond 1\,Mpc, but to obtain
the complete picture of their properties we will also need ultraviolet
(1200-1800\,\AA) spectroscopy of the same stars (or at least a
representative subset). UV spectroscopy provides unique diagnostics of
the intense radiatively-driven winds in massive stars, enabling us to
quantify their velocities, structure, and mass-loss rates (alongside
H$\alpha$). As in visible spectroscopy, we are again at the limit of
current facilities. For instance, UV spectroscopy of individual
massive stars in IC\,1613 (at $\sim$0.7\,Mpc) with the {\em Hubble 
Space Telescope (HST)} required 2 to 5 orbits per star (Garcia et
al. 2014); moving out much beyond 1\,Mpc with {\em HST} is infeasible
within a realistic observing programme.

\begin{figure}
\begin{center}
\includegraphics[height=7cm]{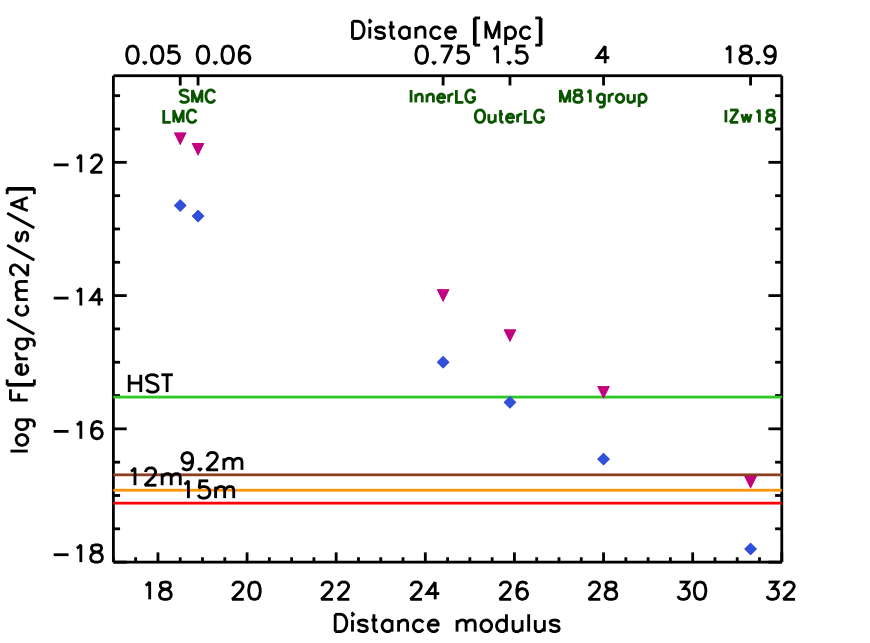} 
\caption{UV fluxes (1500\AA) for O-type dwarfs (diamonds) and B-type
  supergiants (triangles) with increasing distance. The approx.
  sensitivity that can be reached with five {\em HST} orbits is shown
  in green; reaching much beyond 1\,Mpc needs the apertures
  under study for LUVOIR.}\label{fig2}
\end{center}
\end{figure}

Two of the concepts currently under study by NASA in preparation of
the 2020 Decadal Survey could be transformative for the study of
extragalactic massive stars. The {\em Habitable-Exoplanet Imaging
 Mission (HabEx)} concept is studying primary diameters in the range
of 4-6.5m (Gaudi et al. 2018). A multi-object capability on the {\it
 HabEx} UltraViolet Spectrograph (Scowen et al. 2018) would open-up
UV studies of large samples of main-sequence O-type stars in galaxies
beyond 1\,Mpc (e.g. Sextans A) for the first time. Even more exciting
is the study of the {\em Large Ultraviolet/Optical/Infrared Surveyor
 (LUVOIR)} concept, with primaries in the range of 9-15m diameter
(Bolcar et al. 2018). The concept for the LUVOIR Ultraviolet
Multi-Object Spectrograph (LUMOS, France et al. 2017) is particularly
compelling for studies of massive stars. The ultimate goal is for UV
observations of individual stars in I~Zwicky\,18, a very metal
deficient galaxy ($Z$\,$\sim$\,0.02\,$Z_\odot$ at
18.2\,$\pm$\,1.5\,Mpc, Aliosi et al. 2007). To achieve this we
require at least a 12m aperture (see Fig.~\ref{fig2}).

\begin{discussion}
  \discuss{McCarthy}{How critical is GLAO to the performance of the
    MOS instrument?}   
  \discuss{Evans}{It is integral to use of the telescope at these
    wavelengths as the GLAO is needed to provide a stable PSF. As the
    ELT is an adaptive telescope there won't be seeing-limited
    observations in the traditional sense.}
\end{discussion}

\end{document}